\documentstyle[11pt,aaspp4]{article}     

\lefthead{Courteau}
\righthead{AN ATLAS FOR STRUCTURAL STUDIES OF SPIRAL GALAXIES}

\begin{document}



\font\hugemath=cmsy10 scaled \magstep1
\def\Sum{{\hugemath\Sigma}}


\def \etal {{\it et al.~}}
\def \eg {{\it e.g.,~}}
\def \ie {{\it i.e.,~}}
\def \vs {{\it vs~}}
\def\yskip{\penalty-50\vskip3pt plus3pt minus2pt}
\def\pp{\par\yskip\noindent\hangindent 0.4in \hangafter 1}
\def\reference#1#2#3#4 {\pp#1, {\it#2}, {\bf#3}, #4.}


\title{AN ATLAS FOR STRUCTURAL STUDIES OF SPIRAL GALAXIES \
 Rotation Curves and Surface Brightness Profiles of 304 Bright Spirals}

\author{St\'ephane Courteau}

\affil{National Research Council, Herzberg Institute 
       of Astrophysics}
\affil{Dominion Astrophysical Observatory}
\affil{5071 W. Saanich Rd, Victoria, BC \ V8X 4M6 \ Canada}


\begin{abstract}
This is an announcement of a new database of structural properties for 304
late-type (Sb-Sc) spiral galaxies drawn from the UGC catalogue.  These data
were compiled from the kinematic and photometric studies of Courteau (1996,
1997) and are made available to the community via the Canadian Astronomy Data
Centre.  The data base contains redshift information and Tully-Fisher
distances, various measures of optical (H$\alpha$) line width and rotational
velocity, isophotal diameters and magnitudes, disk scale lengths, B-r colour,
rotation curve model parameters, and more.  The main table includes 66 entries
(columns); it can be down-loaded as one single file, or searched for any range
of parameters using our search engine.  The data files for each rotation curve
and luminosity profile (including multiple observations) are also available
and can be retrieved as two separate tar files.  These data were originally
obtained for cosmic flow studies (\eg Courteau \etal 1993, Courteau 1993)
and have been included in the Mark III Catalog of Galaxy Peculiar Velocities
(Willick \etal 1997).  The high spatial and spectral resolution of these
data make them ideal for structural and dynamical investigations of spiral
galaxies (\eg Broeils \& Courteau 1997;  Mo, Mao, \& White 1998; Somerville
\& Primack 1999; Courteau \& Rix 1999).  The data base can be accessed at
{\tt http://cadcwww.hia.nrc.ca/astrocat/courteau.html}.
\end{abstract}




{\noindent {\bf Acknowledgements}}

We would like to thank Daniel Durand at CADC for setting up the Web 
architecture.

\clearpage


\clearpage
\begin{references}
\pp Broeils, A. \& Courteau, S. 1997, {\it Modeling the Mass Distribution 
    in Spiral Galaxies}, in Dark and Visible Matter in Galaxies and Cosmological 
    Implications, ed. Persic \& Salucci, (San Francisco: ASP), Vol. 117, 74

\pp Courteau, S. 1997, {\it Optical Rotation Curves and Linewidths
    for Tully-Fisher Applications}, AJ, 114, 2402 

\pp Courteau, S. 1996, {\it Deep r-band Photometry for Northern Spiral
    Galaxies}, ApJS, 103, 363

\pp Courteau, S. 1993, {\it Comparison of Bulk Flow Solutions},
    in Cosmic Velocity Fields, eds. Bouchet \& Lachi\`eze-Rey
    (\'Editions Fronti\`eres, France), 537

\pp Courteau, S. \& Rix, H-W. 1999, {\it Maximum Disks and the
    Tully-Fisher Relation}, ApJ, 513  

\pp Courteau, S. Faber, S.M., Dressler, A.\& Willick, J.A. 1993, {\it
    Streaming Motions in the Local Universe: Evidence for Large Scale, Low
    Amplitude Density Fluctuations}, ApJL, 412, 51

\pp Mo, H.J., Mao, S., \& White, S.D.M. 1998, 
    {\it The Formation of Galactic Discs}, MNRAS, 295, 319

\pp Somerville, R.S. \& Primack, J.R. 1998, 
    {\it Semi-Analytic Modelling of Galaxy Formation: The Local Universe}, 
    astro-ph/9802268

\pp Willick, J.A., Courteau, S., Faber, S.M., Burstein, D., Dekel, A., \&
    Strauss, M.A. 1997, {\it Homogeneous Velocity-Distance Data for
    Peculiar Velocity Analysis. III.},  
    ApJS, 109, 333


\end{references}
\end{document}